\documentclass[aps, pre, preprint, groupedaddress, amsmath, amssymb]{revtex4-1}
\usepackage{graphicx}  
\usepackage{dcolumn}   
\usepackage{bm}        
\usepackage{multirow}
\begin{document}

\title{A Comparative Study of the Drying Evolution and Dried Morphology of two Globular Protein and De-ionized water Solutions}

\author{Anusuya Pal$^{a}$, Amalesh Gope$^{b}$, Ari S. Athair$^{a}$ and Germano S. Iannacchione$^{a}$} \email{gsiannac@wpi.edu}

\affiliation{$^{a}$ Order-Disorder Phenomena Laboratory, Department of Physics, Worcester Polytechnic Institute, Worcester, MA, 01609, USA \\ $^{b}$ Department of English and Foreign Languages, Tezpur University, Tezpur, Assam, 784028, India}



\begin{abstract}
Pattern formation in drying protein droplets continues to attract considerable research attention because it can be linked to specific protein-protein interactions. An extensive study of the drying evolution and the final crack patterns are presented, highlighting the concentration dependence (from $1$ to $13$~wt\%) on two globular proteins, lysozyme (Lys) and bovine serum albumin (BSA), in de-ionized water. The drying evolution starts with a constant contact radius mode and shifts to a mixed mode where both fluid front and contact angle changes. The contact angle monotonically decreases, whereas, the fluid front exhibits two regimes: an initial linear regime and a later non-linear regime. Unlike the linear regime, the non-linear regime is faster for Lys droplets. This results in the formation of a ``mound"-like structure in the central region. A new feature, a ``dimple" is observed in this mound which is found to be dependent on the initial concentration. The different crack morphology of BSA and Lys depends strongly on the initial state of the solution and can be interpreted using a simple mechanical model. In fact, when dried under uniform conditions (surface, humidity, temperature, droplet diameter, etc.), the evolution and the final pattern displays as a fingerprint of the initial state. 

\end{abstract}


\maketitle


\section*{Introduction}
\label{sec:intro}
A colloidal droplet deposited on a surface either spreads over the surface or remains as it is depending on the wettability of the surface. Whatever be the case, the droplet in general, endures a whole range of interfacial phenomena (wetting dynamics, adsorption, and adhesion), internal flow (diffusion and convection), and particle-substrate interactions during the solvent evaporation (drying) \cite{parsa2018mechanisms, brutin2018recent}. Moreover, the pattern formation of a bio-colloidal droplet, even on an ideal surface, is exposed to an additional complexity during the drying process. This complexity arises primarily due to a strong and potentially competing interparticle interactions that governs the particle aggregation and self-assembly. Furthermore, some bio-colloidal droplets such as blood and plasma serum are ubiquitous, and, most notably, used in medical diagnostics and forensics analysis \cite{brutin2011pattern,chen2017understanding,chen2018controlling,chen2019desiccation}. Studies on drying droplets reveal that the evolution and the emerging patterns depend on multiple factors including the nature of the solute particles (size, chemical composition initial concentration), different type of substrates (hydrophilic, hydrophobic), geometry, substrate wetting, and various drying conditions (temperature, pH, relative humidity) \cite{patil2016effects, dugyala2016role, saenz2017dynamics, retailleau1997no, dunn2009strong}. Accordingly, it turns out to be essential to understand the effects of these factors on the drying and dried bio-colloidal droplets in order to compare (and explain) the macroscopically observed behavior with the initial microscopic state of the constituent particles.

Owing to its significance in many potential applications, the drying evolution and the morphological patterns of dried plasma and blood are examined by several researchers \cite{brutin2011pattern, chen2017understanding,chen2019desiccation}. Essential mechanisms of spreading, gelation, and crack formation of these droplets are presented in the recent review articles by Brutin et al. \cite{smith2018wetting} and Chen et al. \cite{chen2016blood}. An aqueous solution of protein drying droplets may similarly be substantiated as a prolific system of biological relevance. The aqueous solution of proteins is far less complicated than blood and plasma. In a recent study, the drying droplets of raw egg-white protein solutions have been explored that validates the dependence of daisy and wavy-ring crack patterns on the initial protein concentrations \cite{gao2018formation}. 

It is worth mentioning that the commercially available high-quality globular proteins such as bovine serum albumin (BSA) and lysozyme (Lys) have also attracted the attention of many researchers. Many studies investigated the temporal drying process and their resulting patterns. Despite the intense research, most of the work on protein-solvent systems is primarily confined to either a dilute regime of protein concentration and/or the involvement of the salts into the systems \cite{gorr2012lysozyme, gorr2013salt, chen2010complex, annarelli2001crack, carreon2018texture}. The dilute regime of the initial protein concentration is well explored in the drying droplets; however, not beyond the range of $60$~mg/ml. Researchers also attempted to explore the drying evolution and the crack distribution of these proteins dissolved in (de-ionized) water under the ambient conditions \cite{gorr2012lysozyme, carreon2018patterns}. Gorr et al. \cite{gorr2012lysozyme} studied the time evolution and the morphological patterns of Lys dissolved in water varying the initial concentration from $0.1$ ($1$~mg/ml) to $1$~wt\% ($10$~mg/ml). This study concluded that all the drops exhibit a ``coffee-ring" effect \cite{deegan1997capillary, gorr2012lysozyme}. The volume fraction of the lysozyme is found to be linearly dependent on its initial concentration; however, the morphology does not show any significant changes in the ring's height and width. They also reported a ``mound"-like feature in the central region and observed a few surface cracks in the given concentration range. Carre{\'o}n et al. \cite{carreon2018patterns} investigated the mixture of BSA and Lys dissolved in water and mainly focused on their interactions. They also conducted experiments with denatured BSA and Lys proteins and their mixtures at different relative concentrations. The folding and unfolding of these proteins and their structural (morphological) alternation are also discussed in their paper. They concluded that the formation of the crystal clusters and dendrite structures are independent of the external salts. 

In this paper aims to address two fundamental questions about the drying evolution and the resulting patterns of the dried droplets of aqueous solutions of commercially available globular proteins, are addressed. This article aims to explore (i) the role of the protein properties (in terms of mass, composition, configuration, and size); and examine, (ii) the effect of higher initial protein concentration (above $60$~mg/ml and up to $150$~mg/ml) on the aggregation process. The inclusion of the higher-initial protein concentration is essential since the higher concentrations enable us to explore how the excessive aggregation of these proteins play a role in relieving mechanical stress during the drying process, and the crack formation patterns. It is worth mentioning that no studies till date attempted to investigate these fundamental queries at the concentration ranges we considered in this paper. To the best of our knowledge, we did not come across any experimental evidence that compares the drying evolution of the protein droplets in terms of the contact angle and the fluid front. To address this gap, two proteins, BSA and Lys, are chosen in this study. The droplets are prepared using de-ionized water (DI) that avoids ion-mediated effects and exposes the protein-protein interactions. Furthermore, the inclusion of DI under ambient conditions ensures that the functionality of the proteins is not affected by any external factors such as pH, temperature, etc. 

The drying evolution and the emerging patterns are then examined at the initial protein concentration ranging from $1$ ($10$~mg/ml) to $13$~wt\% ($150$~mg/ml). Each solution sample is deposited as a ${\sim}2$~mm diameter droplet on a glass slide to record the temporal variation of contact angle, fluid front, and to capture the image of the final dried state. An image processing technique is employed to extract the distribution of the distance between consecutive cracks (crack spacing) in the protein droplets. To ensure the reliability of the visual observations, an appropriate statistical test is used on the obtained crack-spacing data. A physical mechanism is proposed in this paper to relate the dried morphology with the nature of the initial protein solutions. Finally, the crack patterns are interpreted in terms of a simple mechanical stress model to explain the presence of a crack hierarchy in the Lys (not seen in BSA) droplets.

The paper is structured as follows. Following this introduction, Section~\ref{sec:exp} describes the materials and the experimental methodology adopted in this paper. Section~\ref{sec:results1} presents the process of the drying evolution, the morphology, the physical and mechanical mechanism, and the statistical findings on the dried crack patterns in both protein droplets. The results and the significant findings of this paper are discussed in this section. Finally, the results are concluded in Section~\ref{sec:conc}.

\section{Materials and Experimental Methodology}
\label{sec:exp}

Bovine serum albumin (BSA) and lysozyme (Lys) are well-studied water-soluble globular proteins. BSA is a representative blood protein primarily derived from cows. It is four times heavier than Lys. BSA carries a molecular mass of $ {\sim}66.5$~kDa with an ellipsoid shape of dimensions $4.0 \times 4.0 \times 14.0$~nm$^{3}$ \cite{pal2019phase}. Lys, on the other hand, can be observed in human mucosal secretions such as saliva, tears, etc. It consists of a molecular mass of $ {\sim} 14.3$~kDa, and a roughly ellipsoid shape of dimensions $3.0 \times 3.0 \times 4.5$~nm$^{3}$ \cite{pal2019comparative}. Lys is made up of $129$ amino acids whereas, this blood protein contains $581$ amino acids in a single polypeptide chain. The isoelectric point of BSA is $4.7$ which enables it to carry a net negative charge under the present conditions of the study (pH of ${\sim}7$). The isoelectric point of $11.1$ in Lys, on the other hand, allows it to carry a net positive charge. The globular shape and stability of these proteins are attributed to the disulfide bridges ($17$ in BSA and $4$ in Lys), hydrogen bonds, and hydrophobic interactions among the amino acids \cite{carter1994structure}. It is to be noted that the globular nature of the overall tertiary structure of these proteins is maintained in this present study. The synthetic polymer-based colloidal behavior of these proteins is only possible when these are completely denatured (or their structures are loosed). Therefore, this study is much more complicated and difficult to interpret than other polymer-based droplet studies. 

The commercial lyophilized BSA and the Lys are obtained from Sigma Aldrich, USA (Catalog no. A2153 and L6876 respectively). The protein samples are used without any further purification. ${\sim}150$~mg of each BSA and Lys were massed and separately dissolved in $1$~ml of de-ionized water (Millipore, the resistivity of $18.2$~M$\Omega$.cm) to create the protein stock solutions, BSA+DI and Lys+DI at a concentration of $13$~wt\%. Each stock solution was diluted to prepare the concentrations of $1$, $3$, $5$, $7$, $9$, and $11$~wt\%. To make samples combining protein and liquid crystal (LC), 4-cyano-4'-pentyl-biphenyl was purchased from Sigma Aldrich, USA (Catalog no. 328510). The LC was heated at ${\sim}39$~$^{\circ}$C, and  a  volume of ${\sim}10$~$\mu$L was added to the prepared protein solutions of $9$~wt\%. The sample sets were used in the next few hours of their preparation time. We used the factory-fresh microscopic glass slides as the substrates, which have minimal exposure to the environment prior to the actual experiments. These slides were rinsed with ethanol and dried subsequently. Thus, every droplet precisely contained the same substrate conditions and a uniform reproducibility in terms of the circularity of the droplets and their pinning effects was observed. A volume of $1$~$\pm$~$0.2$~$\mu$L of the sample was pipetted to form a circular droplet of ${\sim}2$~mm diameter. For strip-like geometry, two carbon tapes were placed in parallel on the glass substrate by keeping a separating distance of ${\sim}2$~mm, and a volume of ${\sim}5$~$\mu$L of the protein samples was pipetted. The samples were left to dry under the bright-field microscopy (Leitz Wetzlar, Germany) at a 5$\times$ objective lens in the ambient conditions (room temperature of ${\sim}25$~$^{\circ}$C and relative humidity of ${\sim}50$~\%). The total time of the drying process in all droplets was roughly around $10-20$~minutes. 

The captured images were analyzed using ImageJ \cite{abramoff2004image} software. The time-lapsed images of the drying process were clicked at every two seconds. The start time of the clock was the time of the deposition point of the droplet on the glass slide. The fluid front radius ($r(t)$) was measured by tracking the distance from the center to the edge of the front under the microscopy during the drying evolution. These radii were measured repeatedly ten times at each time ($t$) and the averaged values ($\overline{r}(t)$) were computed. The spiral crack study was conducted using a 50$\times$ objective lens. The images of dried droplets were captured during $24$~hours using side-illuminated bright-field microscopy (since a few cracks appeared after the visible drying process \cite{gorr2013lysozyme}). The \textit{stitching plugin} \cite{preibisch2009globally} of ImageJ was used to redraw the complete image of the dried droplets. The stitched bright-field images were converted into high contrast images. The circular cut lines were drawn using the \textit{oval profile} plugin of ImageJ. The intensity values ($255$ pixels depicts for crack lines and $0$ elsewhere) were plotted as a function of arc-length. A script with  ``Array.andMaxima" was used to determine the positions of the maximum intensity values to estimate the spacing between the consecutive cracks ($x_c$). The detailed processing can be found in our previous paper \cite{pal2019image}. The same process was repeated for two other droplets deposited from the same sample set at each concentration ($\phi$) to ensure the reproducibility of the final morphology. The data were aggregated to yield an overall crack spacing $x_c$ at each $\phi$. It is to be noted that this crack spacing analysis is conducted only at the peripheral region where the cracks of both the protein droplets are present. Furthermore, the contact angle goniometer (Model 90, Ram\'{e}-hart Instrument Co. USA) was used to monitor the contact angle ($\theta(t)$) measurements of the prepared protein samples during the drying process.

\section{Results and Discussions}
\label{sec:results1}
\subsection{Time Evolution of Drying Droplets}

\subsubsection{Bovine Serum Albumin: BSA+DI}

\begin{figure*}[h]
\centering
  \includegraphics[height=7.7cm]{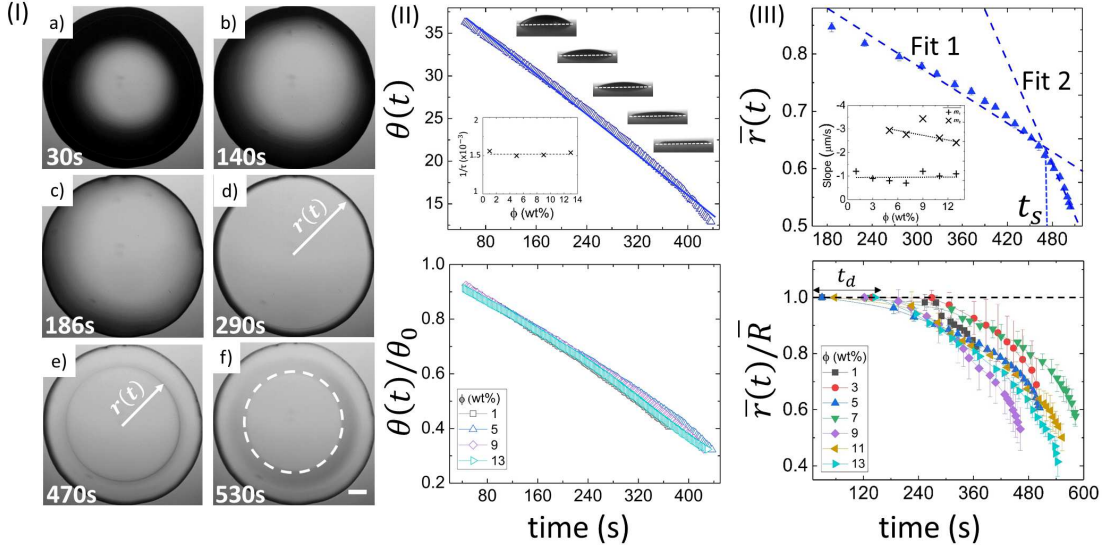}
  \caption{Drying evolution of BSA droplet: (I) Top-view of the droplets through optical microscopy at $\phi$ of $5$~wt\% clicked during different time intervals ($30$~s, $140$~s, $186$~s, $290$~s, $470$~s and $530$~s). The white dashed circle in (f) exhibits the ``coffee-ring" formation. The scale bar represents a length of $0.20$~mm. (II) The top panel shows the variation of contact angle ($\theta(t)$) at $\phi$ of $5$~wt\%. The solid line specifies the fitted function. The inset shows the variation of the characteristic fitting parameter ($1/\tau$) at different $\phi$. The bottom panel displays the temporal variation of the normalized contact angle ($\theta(t)/\theta_{0})$). (III) The top panel reveals the variation of mean fluid front radius ($\overline{r}(t)$) at $\phi$ of $5$~wt\%. The error bars represent the standard deviation. A representative fit of two linear models is made; $t_{s}$ signifies the time point where both the linear fits merge. The inset shows the dependence of the slope values of linear fit 1 ($m_{1}$) and 2 ($m_{2}$). The bottom panel indicates the temporal variation of normalized mean fluid front radius ($\overline{r}(t)/\overline{R}$) at different $\phi$. $t_{d}$ is the ``dead" time up to which the fluid front radius ($r(t)$) is equal to the radius of the droplet ($R$).}
  \label{fig1}
\end{figure*}

Fig.~\ref{fig1}(I) shows the top view of the drying evolution of BSA droplet at the initial concentration ($\phi$) of $5$~wt\% captured through the optical microscopy. As soon as the first image of the deposited droplet is captured, a symmetrical dark (black) shade is observed near the periphery of the droplet. With the progression of time, the dark shade changes to the bright (gray) shade (Fig.~\ref{fig1}(I)a-c). The periphery of the droplet is found to be pinned to the glass substrate throughout the drying process. After ${\sim}3$~minutes, the fluid front starts receding from the periphery to the center of the droplet. The radius ($r$) of this front is measured with the progression of time, ($t$) (Fig.~\ref{fig1}(I)d,e). The movement allows the particles to be deposited along each receding line and eventually forms a peripheral ring (shown by a white dashed circle in Fig.~\ref{fig1}(I)f). It is to be noted that a few cracks in the BSA droplets up to $\phi$ of $5$~wt\% are formed during $24$~hours, which are shown in the following section.

Fig.~\ref{fig1}(II) depicts the side view of the drying evolution of BSA droplet at different $\phi$ observed with the contact angle goniometer. $\theta$ and height at the center of the droplet ($h$) are found to be $37\pm2^{\circ}$ and $0.35\pm0.20$~mm respectively within ${\sim}50$~seconds of the deposition of the droplet. The radius ($R$) of the droplets is $1.1\pm0.2$~mm, and therefore, the gravitational effects can be negligible. Furthermore, the macroscopic shape of the droplet can be approximated as the spherical-cap geometry, which is based on the assumption that $h \ll R$. The black shade, observed at $\phi$ of $5$~wt\% in Fig.~\ref{fig1}(I) is due to the spherical-cap shape of the droplet. The change from the black to the gray shade occurs when the contact angle ($\theta$) reaches to a threshold value. The top panel of Fig.~\ref{fig1}(II) shows the variation of $\theta$ at $\phi$ of $5$~wt\%, and $\theta(t)$ is found to be monotonously reducing with time. This suggests that the drying process occurs in a continuous evaporation limit. To validate this limit, $\theta(t)$ is fitted with a linear function: $\theta (t)=\theta_{0}(1-t/\tau)$, where, $\theta_{0}$ is the initial $\theta$ at $t{=}0$, and $1/\tau$ is a characteristic rate. For $\phi$ of $5$~wt\%, $\theta_{0}$ and $1/\tau$ are found to be $39.96\pm0.08^{\circ}$ and  $0.001500\pm0.000005 s^{-1}$ respectively, with $\text{R}^{2}= 0.995$. The $1/\tau$ is observed to be independent of $\phi$ (inset of Fig.~\ref{fig1}(II)). A complete description of all the fit parameters is tabulated in Table~T1 of the supplementary section. The normalized contact angle is calculated by dividing $\theta(t)$ with the $\theta_{0}$ (obtained from the fitting equation). The individual normalized $\theta$ decay curves at different $\phi$ are found to collapse to a master curve when the data is plotted. This is shown in the bottom panel of Fig.~\ref{fig1}(II). 

In this context, it is interesting to compute the fluid front radius and its dependence on $\phi$. Fig.~\ref{fig1}(III) shows the evolution of the mean fluid front radius ($\overline{r}(t)$) in BSA droplet at different $\phi$. The top panel shows the variation of $\overline{r}(t)$ at $\phi$ of $5$~wt\% exhibiting two distinct regimes: a slow, initial linear regime, and a subsequent non-linear, fast regime. Two linear fits are made on the respective linear and non-linear regimes, and a characteristic time $t_s$ (the time point at which two linear fits intersects) is introduced. It is to be noted that the linearity of $\overline{r}(t)$ deviates after the peripheral ring formation (Fig.~\ref{fig1}(I)e and top panel of Fig.~\ref{fig1}(III)). The inset of Fig.~\ref{fig1}(III) compares the slope values ($m_{1}$ and $m_{2}$) obtained from the linear fits in the respective regimes at each $\phi$. The negative sign in the slope values confirms the reduction of the mean radius ($\overline{r}(t)$) with time. For $1$ and $3$~wt\%, the linear fit in the non-linear regime can't be achieved due to a swift and non-uniform movement, resulting in less number of data points to quantify. On average, the velocity of the fluid front in the linear and non-linear regime is found to be $0.99\pm0.20$~$\mu$m/s and $2.83\pm0.38$~$\mu$m/s respectively. $m_2$ decreases from ${\sim}3$ to ${\sim}2$~$\mu$m/s with the increase of $\phi$. The bottom panel of Fig.~\ref{fig1}(III) displays the normalized radius (obtained by dividing the $\overline{r}(t)$ with the mean radius of the droplet, $\overline{R}$). In the early stage of the drying evolution, i.e., up to ${\sim}240$~seconds, the radius remains constant, ($\overline{r}(t)/\overline{R} = 1$) for all the $\phi$. This time is labeled as the ``dead" time ($t_d$) where only the contact angle changes without disturbing the radius. A complete description of all the measured and fit parameters is tabulated in Table~T2 of the supplementary section.

\subsubsection{Lysozyme: Lys+DI}

\begin{figure*}[t]
\centering
  \includegraphics[height=8.5cm]{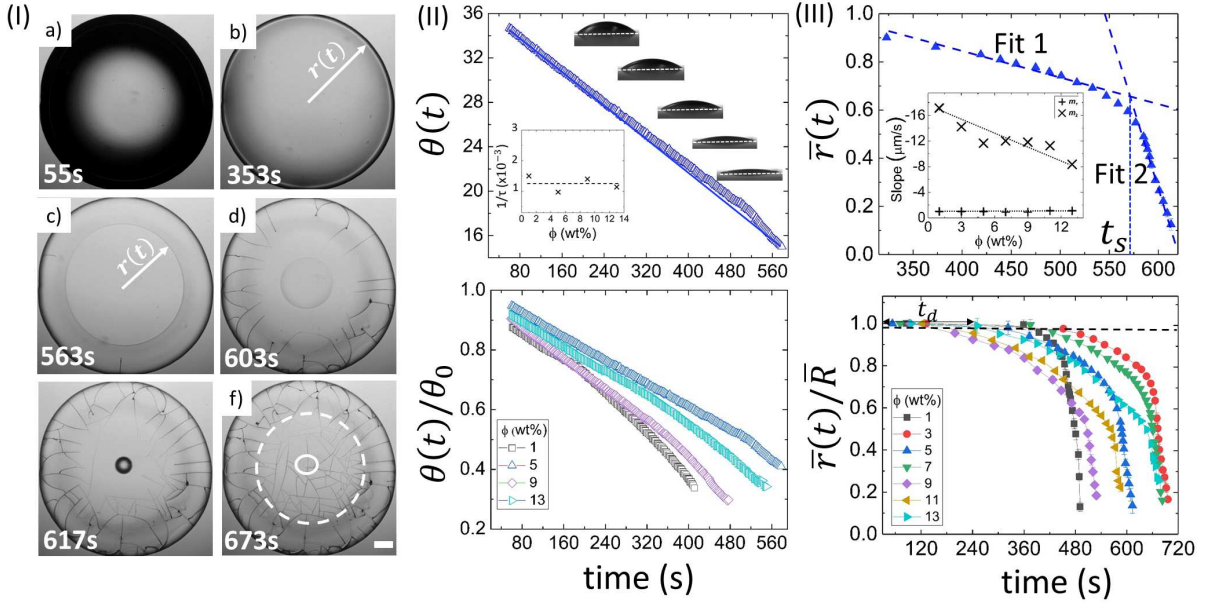}
  \caption{Drying evolution of Lys droplet: (I) Top-view of droplets through optical microscopy at $\phi$ of $5$~wt\% during different drying intervals ($55$~s, $353$~s, $563$~s, $603$~s, $617$~s and $673$~s). The white dashed circle in (f) shows the ``coffee-ring" formation. The solid circle exhibits the ``mound"-like structure. The scale bar represents a length of $0.20$~mm. (II) The top panel confirms a variation of contact angle ($\theta(t)$) at $\phi$ of $5$~wt\%. The solid line shows the fitted function. The inset shows the variation of the characteristic fitting parameter ($1/\tau$) at different $\phi$. The bottom panel displays the temporal variation of the normalized contact angle ($\theta(t)/\theta_{0})$). (III) The top panel reveals the variation of the mean fluid front radius ($\overline{r}(t)$) at $\phi$ of $5$~wt\%. The error bars represent the standard deviation. A representative fit of two linear models is made; $t_{s}$ signifies the time point where both the linear fits merge. The inset represents the dependence of the slope values of linear fit 1 ($m_{1}$) and 2 ($m_{2}$). The bottom panel indicates the temporal variation of the normalized mean fluid front radius ($\overline{r}(t)/\overline{R}$) at different $\phi$. $t_{d}$ is the ``dead" time up to which the fluid front radius ($r(t)$) is equal to the radius of the droplet ($R$).}
  \label{fig2}
\end{figure*}

Akin to BSA, Lys droplet at $\phi$ of $5$~wt\% also shows a dark shade near the periphery of the droplet (Fig.~\ref{fig2}(I)a). The dark shade diminishes, the fluid front starts receding from the periphery after ${\sim}6$~minutes, and forms a ring (Fig.~\ref{fig2}(I)b-d). Interestingly, a sharp spot around the center appears and forms a ``mound"-like structure. The water starts drying from that mound, and finally, a ``dimple" appears in the existing structure. Simultaneously, the radial cracks grow near the periphery and come in each other's contact through the orthoradial cracks (Fig.~\ref{fig2}(I)d-f). The white dashed circle displays the peripheral ring and the solid circle depicts the mound and the dimple structures (Fig.~\ref{fig2}(I)f). Unlike BSA, most of the cracks appeared during the visible drying process at $\phi$ of $5$~wt\%. 

Fig.~\ref{fig2}(II) shows the side view of the drying evolution of Lys droplet at different $\phi$. It is to note that different Lys concentrations might affect the surface tension of the solutions. However, our first measured value of $\theta$ during the contact angle measurements at all $\phi$ is found to be $37.0\pm1.6^{\circ}$. This measurement tempted us conclude that the effect is not significant enough for the unique pattern formation. The $\theta$ at $\phi$ of $5$~wt\% during the drying process reduces monotonously (top panel of Fig.~\ref{fig2}(II)). $\theta_{0}$ and $1/\tau$ are found to be $36.66\pm0.01^{\circ}$ and  $0.0009740\pm0.0000007 s^{-1}$ respectively, with $\text{R}^{2}= 0.999$. Similar to BSA, the characteristic rate ($1/\tau$) is found to be independent of $\phi$. A complete description of all fit parameters is tabulated in Table~T3 of Unlike BSA, the normalized contact angle data shows that the individual decay curves at different $\phi$ start deviating from each other towards the very end of the process (bottom panel of Fig.~\ref{fig2}(II)). 

The top panel of Fig.~\ref{fig2}(III) displays the evolution of normalized mean fluid front radius ($\overline{r}(t)/\overline{R}$) in Lys droplet at different $\phi$. And, the bottom panel depicts the evolution of the $\overline{r}(t)$ at $\phi$ of $5$~wt\%. It is to be noted that this movement in BSA droplets could be tracked only till the point where the radius just passes through the peripheral ring. Unlike BSA, this movement in the Lys droplets could be tracked till the ``mound"-like structure around the central region of the droplet. This causes the range of the $\overline{r}(t)/\overline{R}$ data from $1$ to ${\sim}0.1$. The presence of a linear and a subsequent non-linear regime is commonly observed in the fluid front movement at every $\phi$ in both the Lys and BSA droplets. On average, the velocity of the fluid front in the linear and non-linear regime is found to be $1.00\pm0.08$~$\mu$m/s and $12.36\pm2.73$~$\mu$m/s respectively. A sharp dependence of the slope values in regime~$2$ with $\phi$ is observed, $m_2$ decreases from ${\sim}17$ to ${\sim}8$~$\mu$m/s. A complete description of all the measured and fit parameters is tabulated in Table~T4 of the supplementary section. 

\subsubsection{A Physical Mechanism}

\begin{figure*}[ht]
\centering
  \includegraphics[height=6.3cm]{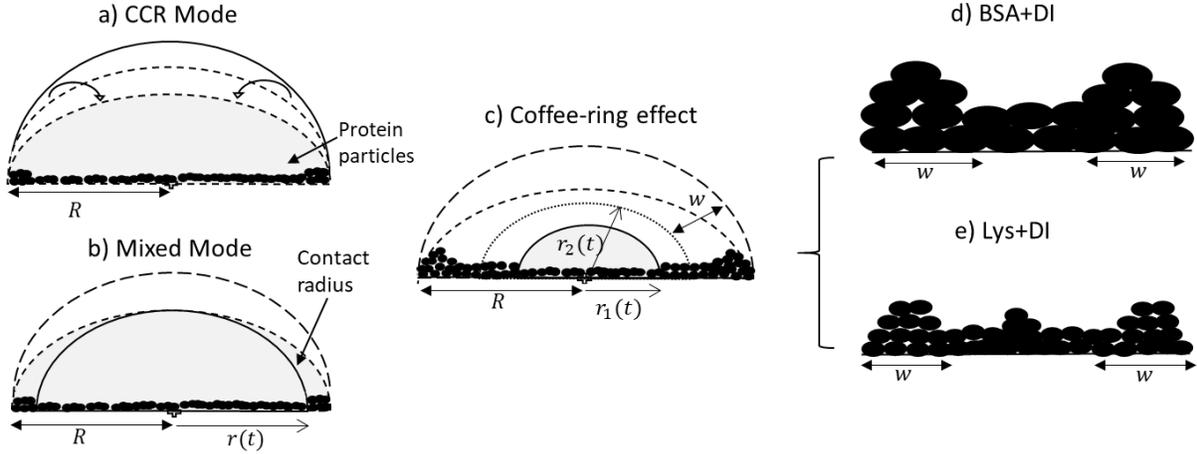}
  \caption{Side view of the drying evolution: a) CCR mode is observed at an early stage of drying where only the contact angle decreases. The fluid front radius ($r$) remains unchanged during this time and equals to the radius ($R$) of the droplet. The protein particles interact with the substrate during this stage. b) In this mixed mode stage, along with the contact angle, the fluid front starts receding from the periphery of the droplet. The fluid front moves, deposits more particles, and forms a ring of width $w$. It moves further when the radius of fluid front just pass the ring width to the new position ($r_{2}$). $r_{1}$ indicates the movement towards the central region of the droplet, leading to a complimentary morphology. d) BSA+DI, without any ``mound"-like structure and e) Lys+DI, the presence of a ``mound" could be observed in the central region.}
  \label{fig3}
\end{figure*}

The underlying physical mechanism of the drying evolution and the visible difference in terms of the morphology of the droplets is demonstrated in Fig.~\ref{fig3}. The deposited droplet goes through a convective flow where the constituent particles tend to interact (adsorb) with the substrate during the early drying stage. The evaporation rate is observed to be highest at the three-phase contact line (solid-vapor-liquid) due to the curvature of the circular droplet. This process drives the flow to compensate the higher rate of mass loss near the periphery compared to the central part of the droplet. This early stage depicted in Fig.~\ref{fig3}a shows the constant contact radius mode (CCR) in which no fluid front movement is observed; however, the height and the contact angle get considerably reduced. During this time frame, the protein particles form an inhomogeneous film and (subsequently) a fluid front starts moving on this film from the periphery to the center which marks the beginning of the next stage. The mixed mode i.e., the movement of both the contact angle and the fluid front is found at this stage. The front seems to deposit some more protein particles as it moves (Fig.~\ref{fig3}b). A bulge (popularly known as ``coffee ring" effect \cite{deegan1997capillary}) at the periphery of the droplet is noticed during this fluid front movement. The bulge is believed to form due to the excess deposition of the particles which could be seen in Fig.~\ref{fig3}c. The drying process till this point, i.e. the formation of this peripheral ring is observed to be similar for both the protein droplets (Figs.~\ref{fig1} and~\ref{fig2}). However, the formation of a ``mound"-like structure (Fig.~\ref{fig3}d,e) in Lys droplets (a similar phenomenon is observed in \cite{carreon2018patterns,gorr2012lysozyme}) creates a visible difference in the drying evolution of both the protein droplets. The comparison of the drying rates of these different sized protein particles may reveal new insights. To map the rate of water loss with the change in morphology during the drying process, it is vital to explore the possible reasons behind the similarities and dissimilarities for both the protein droplets in terms of the parameters ($t_d$, $t_s$, $m_{1}$, and $m_{2}$) extracted from the fluid front movement.

$t_d$ indicates the (early stage) time point of the drying process where the protein particles in both droplets experience the convective flow. During this flow, these particles first tend to interact (adsorb) with the substrate. The usage of de-ionized water in the present study enables us to avoid the ion-mediated effects and does not influence the conformational states (or the functionality) of these proteins. This means that the globular nature of the overall protein structure during the drying process is maintained. During adsorption of these proteins on the glass substrate, one can expect different BSA-glass and Lys-glass interactions. This expectation can be due to the fact that the BSA and the Lys proteins carry an opposite net charge whereas, the glass (substrate) is negatively charged. However, we also need to consider that the hydrophobic residues are buried inside the protein core, and numerous positively and negatively charged residues in a protein's surface are exposed. Though the overall charge of BSA (or Lys) is negative (or positive), it is the fact that the BSA (or Lys) will prefer to adsorb on the negatively charged glass substrate with its positively charged residues. Therefore, the overall interaction of BSA-glass or Lys-glass might not be altered, while there is a high probability of having different BSA-BSA or Lys-Lys interactions. It is because these charged residues help in orientating these protein particles in such a way that one particle gets influenced by the neighboring particle. With time, the protein-protein interactions tempt to be dominant over the protein-substrate interactions, assist in forming the protein film on the substrate, and finally could be responsible for determining these unique patterns. So, $t_d$ is the time in which protein particles interact with the substrate and interact with other proteins to form a film. The constant rate of the evaporation validates of having a similar trend in this CCR mode till the time point $t_d$. 

As time passes during the drying process, we observed a fluid front to recede from the periphery to the center of the droplet. We quantified the velocity of the front movement. The average speed of first linear fit ($m_{1}$) is found to be $0.99\pm0.14$~$\mu$m/s, which is independent of the initial concentration and the type of protein. Considering the trends observed in $m_{1}$, it could be concluded that similar mass transfer mechanisms have emerged in the linear regime. This assumption makes sense because there is enough water on the front surface at this stage, and the front behaves as if it is a water-pool and hardly feels the presence of any protein particles. Subsequently, we have observed a transition from the linear to the non-linear regime in the front movement (Figs.~\ref{fig1}(III) and~\ref{fig2}(III)). A linear fit on the linear and non-linear regimes was made. $t_{s}$ signifies the time point where both these linear fits merge. Interestingly, it can be physically interpreted as the time when the fluid front moves from $r_{2}$ to $r_{1}$ (Fig.~\ref{fig3}c), i.e., the time point of the movement from the edge of the peripheral ring towards the central region of the droplet. The $t_s$ for all $\phi$ is found to be within $\pm20$~s from this ring formation. 

Once the fluid front passes this peripheral ring, the fluid no longer resembles a water-pool. The continuation of water evaporating process leads to the presence of more protein particles than water. In this context, we observed that the velocity of the second linear fit ($m_{2}$) is dependent on the initial concentration and it is different for both the protein droplets. The differences observed in $m_2$ are probably due to different self-assembling interactions, which is dependent on the unique physical characteristics of these proteins. Given the globular nature of these proteins, we know that these proteins are different in terms of their net charged states, molecular shape, weight and disulfide bridges. Since the pH of the system is unchanged, it is beyond the scope of this paper to conclude the mobility effects that emerged due to the individual charged residues present in the protein during the fluid front movement. However, this mobility can easily be interpreted in terms of their weight, shape and bridges. BSA particles are mostly restricted to flow with the fluid front due to high molecular weight (${\sim}66.5$~kDa) and high aspect ratio (major/minor axis = $3.5$). Moreover, the presence of $17$ disulfide bridges in BSA provides a compact network between BSA-BSA particles; it will prefer to be deposited within the existing film-layer in the droplet. This results in a few left-over BSA particles to be carried with the water during the later stage of the fluid front movement. In contrast, Lys could be thought of a squishy sphere (aspect ratio = $1.5$) with a low molecular weight (${\sim}14.3$~kDa). The presence of the lower disulfide bridges ($17$ for BSA and $4$ disulfide bridges for Lys) results in a weak network among these Lys particles, and, it triggers the Lys particles to be carried away with the fluid front. The water content of the fluid decreases with the progression of time, and a large number of Lys particles is left behind. These (left out) particles eventually gets accumulated around the center and forms the ``mound"-like structure. We believe that some water is trapped in the mound. Therefore, a dimple is noticed when the Lys particles fall out of the solution as this entrapped water evaporates. 

The movement of the fluid front appears to slow down with the increase of $\phi$ (number of particles), even though the movement continues to carry and deposit the Lys particles at each line of the fluid front. The concentration dependence of the mound and the peripheral ring on different protein types will be discussed later. At the final stage of the drying process, the water-loss in the droplet induces high mechanical stress leading to the formation of different crack patterns, which will be discussed in the next subsection. The movies of the drying process in both the protein droplets are available in the supplementary videos, V1 for BSA+DI and V2 for Lys+DI; both the videos are recorded at $\phi$ of $5$~wt\%.

\subsection{Morphology of Dried Droplets}

Fig.~\ref{fig4}a-g and Fig.~\ref{fig4}h-n represent the morphology of the dried droplets in BSA and Lys respectively, and both the common and distinctive properties are identified. A few common characteristics of both the dried droplets include (1) the presence of a peripheral ring at a greater height than the central region. This greater height can be viewed from the one-sided dark shadowy shade molded due to side illumination. (2) The cracks are observed in both the droplets; however, the distribution and the nature of the cracks differ to a large extent. 

A few distinct morphological patterns observed in both the protein droplets are (1) In BSA droplets, mostly the radial cracks are observed; whereas, different types of cracks such as radial, wavy, spiral, etc. makes Lys droplets a chaotic system. (2) There are almost no cracks in the central region of the BSA droplet, whereas, the cracks are present throughout the Lys droplets. The cracks are found only in the periphery of the Lys droplets at $\phi$ of $1$~wt\%. (3) There is a complimentary morphology in the central region, with the mound and the dimple structures in Lys and without any such structures in the BSA droplet. (4) The number of cracks increases with the increase of $\phi$ is clearly observed in BSA. The cracks are mostly equally spaced and countable. (5) A thin hair-like structure is observed at the termination of each crack (from $7$ to $13$~wt\%) in BSA droplets. In contrast, the cracks are well-connected and form a uniform domain in every Lys droplet (except the $\phi$ of $1$~wt\%). (6) The presence of (few) circular and (many) spiral cracks in Lys droplets from $9$ to $13$~wt\% makes the morphology very different at the high concentrated regime from that in BSA droplets. The possible reasons are discussed in the mechanical interpretation section.

\subsubsection{Profilometry}

Figs.~\ref{fig5}(I) and (II) show the variation of dimensionless quantity, mean peripheral ring width ($\overline{w}$) divided by the mean radius of the droplet ($\overline{R}$) with $\phi$ for BSA and Lys droplets respectively. It is observed that the ring width is directly proportional to $\phi$, i.e., with the increase of $\phi$, this ring width is expanded further (Fig.~\ref{fig5}(I)). The number of BSA particles increases with the evaporation of water. This process triggers additional deposition of the particles at the droplet periphery with the upsurge of $\phi$. In contrary, the Lys droplet is found to be constant and independent of the variation of $\phi$. An almost equal quantity of Lys particles is deposited in the ring, and most of the free particles are carried towards the center (and forms a mound structure). Fig.~\ref{fig5}(III) shows the areal dependence of this mound structure. The mean area $\overline{a}$ is normalized with $\overline{R}$, and the $\overline{a}/\overline{R}$ at different $\phi$ in Lys droplet is plotted. The linear dependence with $\phi$ makes it evident that most of the Lys particles are carried and deposited towards the center. 

As we have mentioned already, the mound structure is formed by carrying and depositing the free Lys particles during the fluid front movement. As such, the increase of $\phi$ also triggers the upsurge of the overall number of particles, resulting in the piling of these particles to a height. The formation of a dimple (depression) is probably when the Lys particles falls out of the solution. The presence of this dimple is not reported in any of the earlier works \cite{gorr2012lysozyme, carreon2018patterns} probably due to usage of relatively low concentration range. One can anticipate the dimple or the mound as the optical illusion, nonetheless, this is not the case. We did the height profile-like imaging with the sensofar microscopy to confirm the mound and dimple structures; however, we could not calibrate sufficiently to get the exact height measurements. It is also observed that the mound structure is situated almost at the center. It is due to the fact that the circular droplet shape prefers the symmetry for the fluid front movement. To confirm this fact, we pipetted the solutions in a strip-like geometry on the glass substrate. We followed the fluid front movement from both the ends of the strips; however, we did not observe any mound and dimple structures. Furthermore, we assumed that this structure is likely to be shifted to some extent (not forming perfectly at center of the droplet) due to the droplet's circularity. However, no general trend is observed while measuring it in the asymmetrical (or oval) droplets.

\subsubsection{Quantification of the Cracks}

Fig.~\ref{fig6}(I) displays the Q-Q plot at $\phi$ of $7$~wt\% and confirms the non-normal distribution of the mean crack spacing ($x_{c}$). The representative plots also suggest that the cracks are not equally distributed in different protein droplets. The figures indicate that the outliers have surfaced in the form of skewed data points. The outliers (depicted by three circles deviate from the reference line in the Q-Q plot) are not ideal consideration from a statistical perspective since these violate one of the assumptions for t-test (parametric); nonetheless, in our case, there is no good reason to consider these outliers as invalid samples. To counter the non-normal distributions of the mean crack spacing, a non-parametric Mann-Whitney U test (an alternative to the parametric t-test) was preferred to examine the (significant) differences in terms of $x_c$ values among the different protein droplets at different $\phi$. In this study, the mean rank test is chosen over the median (for the visual inspection) because (a) the number of cracks are observed to be different in both the protein droplets; and, (b) the number of samples is relatively large.

In the Mann-Whitney U test, the protein was kept as the categorized factor (independent factor, with two levels, BSA and Lys) and $x_c$ as a dependent variable at different $\phi$. All the histograms are expressed as the mean $\pm$ standard error (SE). The differences where p $\leq 0.05$ are considered to be significant in this study. A detailed report of the statistical test (U, z, and p values) is shown in Table~T5 of the supplementary section. Fig.~\ref{fig6}(II) shows $\overline{x}_c$ for BSA and Lys droplets at each $\phi$. The results confirm our morphological observations that the crack patterns in BSA and Lys droplets are different from each other, resulting in a significant difference in $ \overline{x}_c$ in the peripheral regions at every $\phi$ (the presence of crack spacing at every  $\phi$ is significantly higher in Lys). The visual observations could be considered as an effective way to determine the differences in terms of $\overline{x}_c$; however, the statistical test helps us to distinguish the data fluctuations borne out due to the experimental conditions. It is observed that $\overline{x}_c$ in the Lys droplets varies from $0.08$ to $0.15$~mm without any trend when the $\phi$ is increased. In contrast, $ \overline{x}_c$ in the BSA droplet varies from $0.19$ to $0.27$~mm, and suggests a proportionate distribution; i.e., $x_c$ increase with the increased $\phi$. The release of the available stress is mostly unidirectional in the BSA droplets, resulting in a uniform crack pattern in the peripheral ring. The uniform crack spacing resulted in an increased $x_c$ when $\phi$ is also increased. However, in Lys droplets, the stress is relieved from all the directions resulting in the distribution of small to large cracks spread throughout the droplet, affecting $\overline{x}_c$ at each $\phi$.

Fig.~\ref{fig6}(III) shows a characteristic linear fit of a spiral crack trajectory at $\phi$ of $13$~wt\% in the Lys droplet. The domains containing the spiral cracks in the Lys droplets are three dimensional and it is not possible to observe all the spiral lines simultaneously with the droplet surface. The lack of information about the $z$ plane prompted us to represent these spirals on the $xy$ plane (in the form of 2D). The expression of the logarithmic spiral in polar coordinates: $s(\theta) = a e^{b\theta}$. Assumption of this logarithm leads to $\ln s(\theta) = \ln a + b \theta$ where, $s(\theta)$ is the distance from the spiral center, and $\theta $ is the angle which is in an anti-clockwise direction from the $x$ axis, not restricted to 2$\pi$. The final theta corresponds to the angle made by the $x$ axis and the outermost spiral line. The schematic diagram of a two-dimensional spiral is shown in the upper inset. The logarithmic parameters are ``a", which is the apparent length of the spiral, and ``b", which controls the tightness and predicts the direction of the spiral. A lower value of ``b" means the spiral has more revolutions and hence, more tightness in the spiral shape. No preference of clockwise or counter-clockwise direction in the spirals is observed in any of these droplets. This prompted us to generalize the direction by flipping all the required images so that the spirals would be consistent every time with the starting spiral revolution line lying at zero degrees as shown in the lower inset. The linear fit between $\ln s(\theta)$ and $\theta$ with $ \text{R}^{2} $ of $0.957$ confirms that the spirals in Lys droplets are in the form of logarithmic spirals. An oscillation of the data points is obtained due to the presence of the irregular, polygonal-shaped domains. The overall shape of the spirals for different $ \phi $ is almost the same; however, the trajectories are influenced by the material and fracture properties (a similar phenomenon is observed in \cite{neda2002spiral,sendova2003spiral}). The spirals at $ \phi $ of $13$~wt\% are shown in Fig.~S1 of the supplementary section, and various parameters of the spiral crack analysis at $\phi$ of $11$ and $13$~wt\% are reported in Table~T6 of the supplementary section. Spirals with a very few revolutions in $ \phi $ of $13$~wt\% are also observed. The value of ``b" is found to be in the range of $0.0376$ to $0.0548$~$\mu$m/rad- a narrow range implying that the tightness is probably insensitive to the concentration of proteins; however, a detailed trend of ``b" requires to examine more levels $ \phi $.

\subsubsection{A Mechanical Interpretation}

It is reported in the earlier subsection that each droplet is pinned to the substrate throughout the drying process. The particles in the droplet are adsorbed on the substrate, and simultaneously are carried towards the periphery. With further water-loss from the droplet, the protein particles are deposited in such a way that it creates a film. These particles are accumulated in the layers and might be influenced by a shear-mode or mode II (the stress is applied parallel to the plane). However, this influence is almost negligible as the top surface of the film still contains enough water to evaporate. This water evaporates during the fluid front movement, and the tensile stressed fields are generated when the droplet is almost devoid of water. 

Fig.~\ref{fig7} indicates that two types of tensile stresses (mode I) are involved in propagating the radial and azimuthal cracks in the protein droplets. $\sigma_{\theta}$ and $\sigma_{r}$ are the stress that normally acts to the radial crack and azimuthal crack, respectively. In both the droplets, a directional growth, i.e., a radial crack, was initially observed to propagate from the periphery of the droplet (for example, see Fig.~\ref{fig2}(I)c). Therefore, it indicates that the stress acts along the fluid front, normal to the radial crack, i.e., $\sigma_{\theta}$ is dominant initially. It is known from Griffith's hypothesis that the moment when the available stress in the film exceeds the critical stress, the excess stress is released by virtue of the crack propagation \cite{katzav2007theory,goehring2015desiccation}. This film height could be one of the reasons for the cracks appearing in the peripheral ring first and then proceeds towards the central region in every protein droplet in general. We attempted the height profile-like imaging with the sensofar microscopy; however, we could not calibrate sufficiently to get the exact height measurements. The crack propagation could also be characterized based on the opening of the cracks and the distance of the crack tip as a function of $\phi$. However, this is not possible with the current set-up as the time-lapse images are captured with an 8-bit camera and can only be taken every two seconds. 

\begin{figure*}[t]
\centering
  \includegraphics[height=5.0cm]{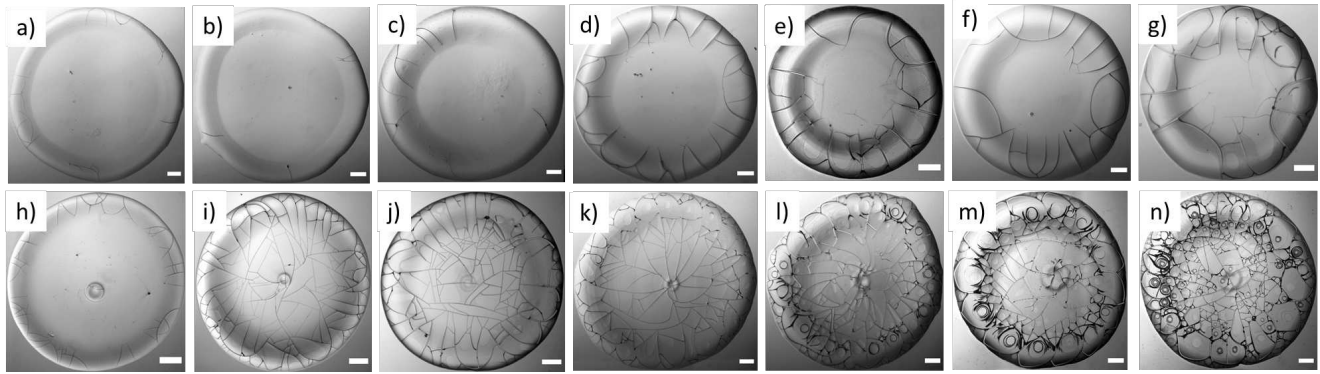}
  \caption{Gray images of two protein dried droplets: BSA at $\phi$ of a) $1$, b) $3$, c) $5$, d) $7$, e) $9$, f) $11$, and g) $13$~wt\%. Lys at $\phi$ of h) $1$, i) $3$, j) $5$, k) $7$, l) $9$, m) $11$, and n) $13$~wt\%. The scale bar represents a length of $0.20$~mm. The images of the BSA droplets reveal the presence of the cracks near the periphery region, whereas, the images of Lys droplets confirm that the cracks are spread and connected from $\phi$ of $3$~wt\%. The ``dimple" and the ``mound"-structure can be observed in Lys droplets at every $\phi$. The spirals are noticed in Lys droplets from $\phi$ of $9$~wt\%.}
  \label{fig4}
\end{figure*}

\begin{figure*}[ht]
 \centering
 \includegraphics[height=4.5cm]{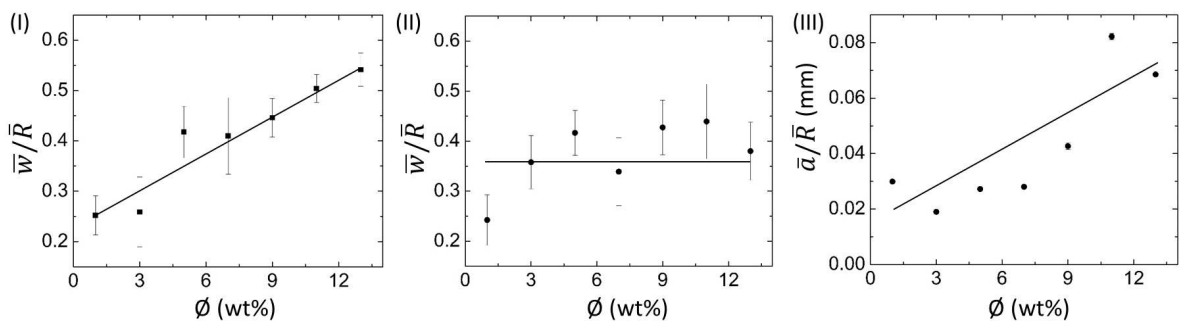}
 \caption{Profilometric measurements of protein droplets: Variation of the mean peripheral ring width ($\overline{w}$) normalized to the mean radius of the droplet ($\overline{R}$) with $\phi$ for (I) BSA+DI, and (II) Lys+DI. (III) Variation of the mean area of ``mound"-like structure ($\overline{a}$) normalized to $\overline{R}$ with $\phi$ in Lys droplet. The error bars correspond to the standard deviation obtained from multiple measurements.}
 \label{fig5}
\end{figure*}

\begin{figure*}[ht]
\centering
  \includegraphics[height=4.0cm]{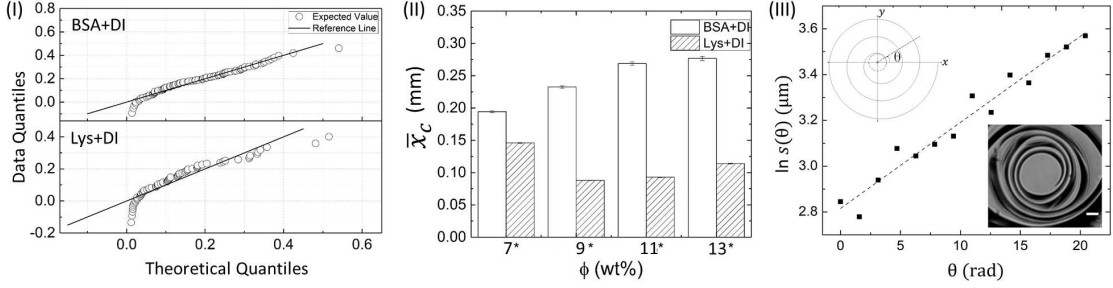}
  \caption{I) Q-Q plot of BSA and Lys droplet at $\phi$  of $7$~wt\% displaying the skewed data points. (II) The histogram depicting the comparison of the mean crack spacing ($\overline{x}_c$) at different $\phi$ among the proteins in the peripheral regions. Significant pairs (BSA and Lys) are marked with an asterisk [*] at each $\phi$. The error bars correspond to the standard error. (III) A characteristic linear fit of a spiral crack trajectory at $\phi$ of $13$~wt\% in Lys droplet. The upper inset shows the schematic diagram of the spiral cracks projected on the $xy$ plane, in which $s(\theta)$ is the radial distance from the spiral center and $\theta $ is the angle starting from zero in an anticlockwise direction. The lower inset shows an example of spiral focused at $50\times$ objective lens, with the scale bar of length $10$~$\mu $m.}
  \label{fig6}
\end{figure*}

\begin{figure*}[ht]
\centering
  \includegraphics[height=4.5cm]{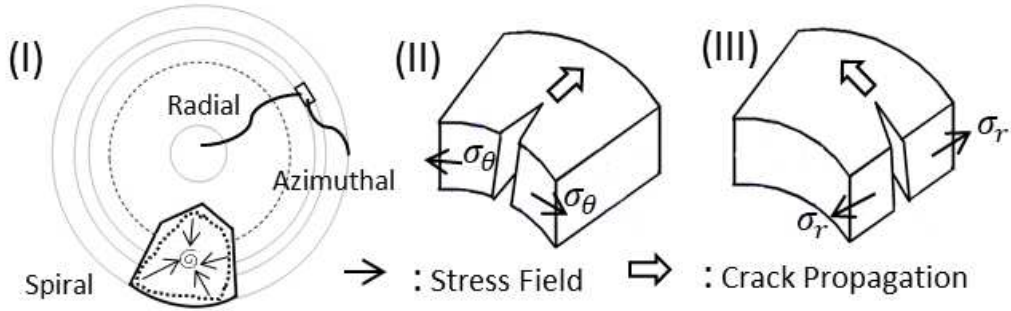}
  \caption{Schematic illustration explaining of the nature and propagation of cracks in both the protein droplets. (I) Top-view of a droplet shows radial and azimuthal cracks. A spiral crack is also shown in a polygonal crack domain. An element is focused on displaying crack propagation to get a notion of stress. Mode I (tensile mode) in which (II) stress ($\sigma_{\theta}$) acting normal to the radial crack and (III) stress ($\sigma_{r}$) acting normal to the azimuthal crack.}
  \label{fig7}
\end{figure*}

After the formation of the radial cracks, a few are curved azimuthally and the remaining ones join the neighboring cracks. However, it is observed that the stressed fields are dependent on the nature of protein particles. The crack propagation is stopped in BSA droplets, and a hair-like crack is developed without invading the central region of the droplet (Fig.~\ref{fig4}d-g). The cracks form almost no connected domains. Our assumption is that the propagation of the cracks is stopped when the film thickness is less than the critical crack thickness resulting in zero cracks in the central region of every BSA droplet.

Lys, on the other hand, is a loosely-compacted and a low molecular weighted protein. The cracks are spread all over the droplet at all $\phi$ except $1$~wt\%. The presence of mound and dimple structures in the central region creates a thick gradient (highest at the periphery, lower at the center, and lowest at the middle region) during the drying process. This thickness gradient is enough to meet the crack propagation criteria and joins the crack lines from the center to the periphery. The stress fields act from all the directions ($\sigma_{r}$ and $\sigma_{\theta}$) leading to a chaotic system. No cracks are observed in the central region at $1$~wt\% (Fig.~\ref{fig4}h) due to the presence of a few Lys particles in the middle region, which reduces the film thickness from the critical crack thickness. There are mostly well-connected polygonal domains in the Lys droplets at $\phi$ of $3$~wt\% and above (Fig.~\ref{fig4}i-n). The cracking leads to a subsequent process of delamination at $\phi$ of $5$~wt\% and $7$~wt\% near the periphery of the droplet (Fig.~\ref{fig4}j,k). Adhesion energy persists between the protein particles and the glass (substrate). As soon as the stored elastic energy in the domain overcomes this adhesion energy, each domain buckled- curving outwards like a bowl (a similar process is observed in other studies as well \cite{sobac2014desiccation, giorgiutti2015striped,lazarus2011craquelures}). The interference fringes of each fragmented domain imply that there is an air gap between the detached film and the substrate that forms a non-uniform adhering region in each domain present in the Lys droplet. This phenomenon is observed in high-magnification images of Lys droplets represented in Fig.~\ref{fig4}j-n. 

A spiral path, thus, is initiated in the well-connected polygonal domain of the Lys droplets at $\phi$ from $9$~wt\% onwards (Fig.~\ref{fig4}l-n). The spirals propagate to release the elastic energy stored in the fragmented domain. This phenomenon is observed from $\phi$ of $9$~wt\% in Lys droplet, implying that there is not enough stored elastic energy in the fragmented domains below $\phi$ of $9$~wt\%. The formation of the spirals on the irregular (polygonal) domains in Lys droplets has no radial cracks intersecting the spiral cracks, i.e., no splitting of the spiral cracks is observed in the Lys droplet (Fig.~\ref{fig4}m,n). It indicates that the size domains turn so small that there is no available energy for the radial cracks to propagate. At $\phi$ of $9$ and $11$~wt\% in Lys droplets, it is observed that the corners of the domains act as the precursor (Fig.~\ref{fig4}l,m). It means that a high-stress area is achieved; however, due to lack of sufficient energy, the spirals fail to form the shape of those observed at $\phi$ of $13$~wt\% (Fig.~\ref{fig4}n). Most of the spirals are present on the peripheral ring of the droplet at $\phi$ of $11$ and $13$~wt\% (Figs.~\ref{fig4}m,n). This implies that the film height could be one of the necessary criteria to have spiral cracks only at specific $\phi$ in the Lys droplet. 

Though the increase of the protein concentration directly increases the film thickness, the heavy weighted protein (BSA of ${\sim}66.5$~kDa) contains less number of particles forming thinner film height than that observed in Lys (${\sim}14.3$~kDa) at the same initial concentration. We argue that the absence of well-connected domains makes it hard to buckle the protein film. As a result, the stored elastic energy can not be applied from all the directions on the delamination front to propagate in the BSA droplet. This is evident when we studied these proteins at $\phi$ of $9$~wt$\%$ in the presence of $5$CB liquid crystal (LC). None of the LC was present in the crack lines, and all were carried underneath the domains in Lys. The center of each domain adhered to the substrate and therefore appeared black under crossed polarizing configuration \cite{pal2019phase}. In contrast, some of the LC was found in the crack lines and mostly distributed on the top of the BSA film \cite{pal2019comparative}. These studies indicate that the domains of the BSA adhered to the substrate, but, Lys did not do so. 

Interestingly, a hierarchy is only observed in the Lys protein droplets. At the lowest $\phi$ ($1$~wt\%), the cracks are present in the peripheral ring of Lys droplet. The cracks are observed throughout the film from $3$ to $13$~wt\%; however, there is no delamination process involved until the $\phi$ of $3$~wt\%. At $\phi$ of $5$ and $7$~wt\%, the delaminated cracks are observed, particularly in the ring. As $\phi$ increased from $9$ to $13$~wt\%, the circular and spiral cracks appeared in the ring in addition to the delamination (Fig.~\ref{fig4}h-n). Observation of these spirals might be a common phenomenon in polymeric systems \cite{dillard1994spiral,behnia2017spiral,neda2002spiral}; however, however, such phenomenon observed in Lys droplets is yet to be reported. This proves a similar unstructured, amorphous reminiscent behavior of the Lys protein. 

This mechanical interpretation, thus, reveals the differences in the type of crack patterns observed in the BSA and Lys droplets and throws light on the reason behind the existence of spirals in the Lys droplets at specific $\phi$. Further, this mechanical interpretation can also be used in explaining the crack patterns in the dried droplets of any bio-molecules.

\section{Conclusions}
\label{sec:conc}

This work showcased the self-assembly of proteins and demonstrates that the process of self-assembly is driven by the drying process. The findings of the experiments confirm that the nature of protein plays an important role in deciding the drying evolution and the subsequent morphology. The monotonous reduction of the contact angle during the drying process helps in identifying different modes of the evaporation. The relatively higher initial protein concentration used in this study facilitated in identifying a ``dimple" on the mound-structure in the dried lysozyme droplets. This study further establishes the presence of a spiral crack pattern at the specific initial protein concentration in lysozyme droplets, which has not been arrested in the literature so far. The non-parametric statistical tests facilitate the crack spacing quantification and confirm the visual observations. This procedure of quantification may be used in broad disciplines to quantify different parameters and their effects. 

It is to be noted that all the experiments in this paper have been performed in DI water, and the presence of ions in body fluids will influence the patterns to a great extent. However, this study can set a baseline for understanding the multi-component systems such as proteins with the addition of various salts (ions), whole human blood, plasma serum, etc. when dried under uniform conditions (surface, humidity, temperature, droplet diameter, etc.). The resulting pattern of the drying droplets is expected to be a signature of the initial state, as observed in our study. Thus, these findings of this paper ensure that such information may potentially to be used for several diagnostic screening in the near future.

\section*{Acknowledgments}
This work is supported by the Department of Physics at WPI. The authors would also like to thank the Tinkerbox (Innovation and Entrepreneurship) Grant sponsored by Women Impact Network (WIN) at WPI for the financial assistance and moral support.


\bibliography{comref}

\begin{thebibliography}{37}%
\makeatletter
\providecommand \@ifxundefined [1]{%
 \@ifx{#1\undefined}
}%
\providecommand \@ifnum [1]{%
 \ifnum #1\expandafter \@firstoftwo
 \else \expandafter \@secondoftwo
 \fi
}%
\providecommand \@ifx [1]{%
 \ifx #1\expandafter \@firstoftwo
 \else \expandafter \@secondoftwo
 \fi
}%
\providecommand \natexlab [1]{#1}%
\providecommand \enquote  [1]{``#1''}%
\providecommand \bibnamefont  [1]{#1}%
\providecommand \bibfnamefont [1]{#1}%
\providecommand \citenamefont [1]{#1}%
\providecommand \href@noop [0]{\@secondoftwo}%
\providecommand \href [0]{\begingroup \@sanitize@url \@href}%
\providecommand \@href[1]{\@@startlink{#1}\@@href}%
\providecommand \@@href[1]{\endgroup#1\@@endlink}%
\providecommand \@sanitize@url [0]{\catcode `\\12\catcode `\$12\catcode
  `\&12\catcode `\#12\catcode `\^12\catcode `\_12\catcode `\%12\relax}%
\providecommand \@@startlink[1]{}%
\providecommand \@@endlink[0]{}%
\providecommand \url  [0]{\begingroup\@sanitize@url \@url }%
\providecommand \@url [1]{\endgroup\@href {#1}{\urlprefix }}%
\providecommand \urlprefix  [0]{URL }%
\providecommand \Eprint [0]{\href }%
\providecommand \doibase [0]{http://dx.doi.org/}%
\providecommand \selectlanguage [0]{\@gobble}%
\providecommand \bibinfo  [0]{\@secondoftwo}%
\providecommand \bibfield  [0]{\@secondoftwo}%
\providecommand \translation [1]{[#1]}%
\providecommand \BibitemOpen [0]{}%
\providecommand \bibitemStop [0]{}%
\providecommand \bibitemNoStop [0]{.\EOS\space}%
\providecommand \EOS [0]{\spacefactor3000\relax}%
\providecommand \BibitemShut  [1]{\csname bibitem#1\endcsname}%
\let\auto@bib@innerbib\@empty
\bibitem [{\citenamefont {Parsa}\ \emph {et~al.}(2018)\citenamefont {Parsa},
  \citenamefont {Harmand},\ and\ \citenamefont
  {Sefiane}}]{parsa2018mechanisms}%
  \BibitemOpen
  \bibfield  {author} {\bibinfo {author} {\bibfnamefont {M.}~\bibnamefont
  {Parsa}}, \bibinfo {author} {\bibfnamefont {S.}~\bibnamefont {Harmand}}, \
  and\ \bibinfo {author} {\bibfnamefont {K.}~\bibnamefont {Sefiane}},\
  }\href@noop {} {\bibfield  {journal} {\bibinfo  {journal} {Advances in
  colloid and interface science}\ }\textbf {\bibinfo {volume} {254}},\ \bibinfo
  {pages} {22} (\bibinfo {year} {2018})}\BibitemShut {NoStop}%
\bibitem [{\citenamefont {Brutin}\ and\ \citenamefont
  {Starov}(2018)}]{brutin2018recent}%
  \BibitemOpen
  \bibfield  {author} {\bibinfo {author} {\bibfnamefont {D.}~\bibnamefont
  {Brutin}}\ and\ \bibinfo {author} {\bibfnamefont {V.}~\bibnamefont
  {Starov}},\ }\href@noop {} {\bibfield  {journal} {\bibinfo  {journal}
  {Chemical Society Reviews}\ }\textbf {\bibinfo {volume} {47}},\ \bibinfo
  {pages} {558} (\bibinfo {year} {2018})}\BibitemShut {NoStop}%
\bibitem [{\citenamefont {Brutin}\ \emph {et~al.}(2011)\citenamefont {Brutin},
  \citenamefont {Sobac}, \citenamefont {Loquet},\ and\ \citenamefont
  {Sampol}}]{brutin2011pattern}%
  \BibitemOpen
  \bibfield  {author} {\bibinfo {author} {\bibfnamefont {D.}~\bibnamefont
  {Brutin}}, \bibinfo {author} {\bibfnamefont {B.}~\bibnamefont {Sobac}},
  \bibinfo {author} {\bibfnamefont {B.}~\bibnamefont {Loquet}}, \ and\ \bibinfo
  {author} {\bibfnamefont {J.}~\bibnamefont {Sampol}},\ }\href@noop {}
  {\bibfield  {journal} {\bibinfo  {journal} {Journal of Fluid Mechanics}\
  }\textbf {\bibinfo {volume} {667}},\ \bibinfo {pages} {85} (\bibinfo {year}
  {2011})}\BibitemShut {NoStop}%
\bibitem [{\citenamefont {Chen}\ \emph {et~al.}(2017)\citenamefont {Chen},
  \citenamefont {Zhang}, \citenamefont {Zang},\ and\ \citenamefont
  {Shen}}]{chen2017understanding}%
  \BibitemOpen
  \bibfield  {author} {\bibinfo {author} {\bibfnamefont {R.}~\bibnamefont
  {Chen}}, \bibinfo {author} {\bibfnamefont {L.}~\bibnamefont {Zhang}},
  \bibinfo {author} {\bibfnamefont {D.}~\bibnamefont {Zang}}, \ and\ \bibinfo
  {author} {\bibfnamefont {W.}~\bibnamefont {Shen}},\ }\href@noop {} {\bibfield
   {journal} {\bibinfo  {journal} {Journal of Materials Chemistry B}\ }\textbf
  {\bibinfo {volume} {5}},\ \bibinfo {pages} {8991} (\bibinfo {year}
  {2017})}\BibitemShut {NoStop}%
\bibitem [{\citenamefont {Chen}\ \emph {et~al.}(2018)\citenamefont {Chen},
  \citenamefont {Zhang},\ and\ \citenamefont {Shen}}]{chen2018controlling}%
  \BibitemOpen
  \bibfield  {author} {\bibinfo {author} {\bibfnamefont {R.}~\bibnamefont
  {Chen}}, \bibinfo {author} {\bibfnamefont {L.}~\bibnamefont {Zhang}}, \ and\
  \bibinfo {author} {\bibfnamefont {W.}~\bibnamefont {Shen}},\ }\href@noop {}
  {\bibfield  {journal} {\bibinfo  {journal} {Journal of Materials Chemistry
  B}\ }\textbf {\bibinfo {volume} {6}},\ \bibinfo {pages} {5867} (\bibinfo
  {year} {2018})}\BibitemShut {NoStop}%
\bibitem [{\citenamefont {Chen}\ \emph {et~al.}(2019)\citenamefont {Chen},
  \citenamefont {Zhang}, \citenamefont {He},\ and\ \citenamefont
  {Shen}}]{chen2019desiccation}%
  \BibitemOpen
  \bibfield  {author} {\bibinfo {author} {\bibfnamefont {R.}~\bibnamefont
  {Chen}}, \bibinfo {author} {\bibfnamefont {L.}~\bibnamefont {Zhang}},
  \bibinfo {author} {\bibfnamefont {H.}~\bibnamefont {He}}, \ and\ \bibinfo
  {author} {\bibfnamefont {W.}~\bibnamefont {Shen}},\ }\href@noop {} {\bibfield
   {journal} {\bibinfo  {journal} {ACS sensors}\ }\textbf {\bibinfo {volume}
  {4}},\ \bibinfo {pages} {1701} (\bibinfo {year} {2019})}\BibitemShut
  {NoStop}%
\bibitem [{\citenamefont {Patil}\ \emph {et~al.}(2016)\citenamefont {Patil},
  \citenamefont {Bange}, \citenamefont {Bhardwaj},\ and\ \citenamefont
  {Sharma}}]{patil2016effects}%
  \BibitemOpen
  \bibfield  {author} {\bibinfo {author} {\bibfnamefont {N.~D.}\ \bibnamefont
  {Patil}}, \bibinfo {author} {\bibfnamefont {P.~G.}\ \bibnamefont {Bange}},
  \bibinfo {author} {\bibfnamefont {R.}~\bibnamefont {Bhardwaj}}, \ and\
  \bibinfo {author} {\bibfnamefont {A.}~\bibnamefont {Sharma}},\ }\href@noop {}
  {\bibfield  {journal} {\bibinfo  {journal} {Langmuir}\ }\textbf {\bibinfo
  {volume} {32}},\ \bibinfo {pages} {11958} (\bibinfo {year}
  {2016})}\BibitemShut {NoStop}%
\bibitem [{\citenamefont {Dugyala}\ \emph {et~al.}(2016)\citenamefont
  {Dugyala}, \citenamefont {Lama}, \citenamefont {Satapathy},\ and\
  \citenamefont {Basavaraj}}]{dugyala2016role}%
  \BibitemOpen
  \bibfield  {author} {\bibinfo {author} {\bibfnamefont {V.~R.}\ \bibnamefont
  {Dugyala}}, \bibinfo {author} {\bibfnamefont {H.}~\bibnamefont {Lama}},
  \bibinfo {author} {\bibfnamefont {D.~K.}\ \bibnamefont {Satapathy}}, \ and\
  \bibinfo {author} {\bibfnamefont {M.~G.}\ \bibnamefont {Basavaraj}},\
  }\href@noop {} {\bibfield  {journal} {\bibinfo  {journal} {Scientific
  Reports}\ }\textbf {\bibinfo {volume} {6}},\ \bibinfo {pages} {30708}
  (\bibinfo {year} {2016})}\BibitemShut {NoStop}%
\bibitem [{\citenamefont {S{\'a}enz}\ \emph {et~al.}(2017)\citenamefont
  {S{\'a}enz}, \citenamefont {Wray}, \citenamefont {Che}, \citenamefont
  {Matar}, \citenamefont {Valluri}, \citenamefont {Kim},\ and\ \citenamefont
  {Sefiane}}]{saenz2017dynamics}%
  \BibitemOpen
  \bibfield  {author} {\bibinfo {author} {\bibfnamefont {P.}~\bibnamefont
  {S{\'a}enz}}, \bibinfo {author} {\bibfnamefont {A.}~\bibnamefont {Wray}},
  \bibinfo {author} {\bibfnamefont {Z.}~\bibnamefont {Che}}, \bibinfo {author}
  {\bibfnamefont {O.}~\bibnamefont {Matar}}, \bibinfo {author} {\bibfnamefont
  {P.}~\bibnamefont {Valluri}}, \bibinfo {author} {\bibfnamefont
  {J.}~\bibnamefont {Kim}}, \ and\ \bibinfo {author} {\bibfnamefont
  {K.}~\bibnamefont {Sefiane}},\ }\href@noop {} {\bibfield  {journal} {\bibinfo
   {journal} {Nature Communications}\ }\textbf {\bibinfo {volume} {8}},\
  \bibinfo {pages} {14783} (\bibinfo {year} {2017})}\BibitemShut {NoStop}%
\bibitem [{\citenamefont {Retailleau}\ \emph {et~al.}(1997)\citenamefont
  {Retailleau}, \citenamefont {Ries-Kautt},\ and\ \citenamefont
  {Ducruix}}]{retailleau1997no}%
  \BibitemOpen
  \bibfield  {author} {\bibinfo {author} {\bibfnamefont {P.}~\bibnamefont
  {Retailleau}}, \bibinfo {author} {\bibfnamefont {M.}~\bibnamefont
  {Ries-Kautt}}, \ and\ \bibinfo {author} {\bibfnamefont {A.}~\bibnamefont
  {Ducruix}},\ }\href@noop {} {\bibfield  {journal} {\bibinfo  {journal}
  {Biophysical Journal}\ }\textbf {\bibinfo {volume} {73}},\ \bibinfo {pages}
  {2156} (\bibinfo {year} {1997})}\BibitemShut {NoStop}%
\bibitem [{\citenamefont {Dunn}\ \emph {et~al.}(2009)\citenamefont {Dunn},
  \citenamefont {Wilson}, \citenamefont {Duffy}, \citenamefont {David},\ and\
  \citenamefont {Sefiane}}]{dunn2009strong}%
  \BibitemOpen
  \bibfield  {author} {\bibinfo {author} {\bibfnamefont {G.}~\bibnamefont
  {Dunn}}, \bibinfo {author} {\bibfnamefont {S.}~\bibnamefont {Wilson}},
  \bibinfo {author} {\bibfnamefont {B.}~\bibnamefont {Duffy}}, \bibinfo
  {author} {\bibfnamefont {S.}~\bibnamefont {David}}, \ and\ \bibinfo {author}
  {\bibfnamefont {K.}~\bibnamefont {Sefiane}},\ }\href@noop {} {\bibfield
  {journal} {\bibinfo  {journal} {Journal of Fluid Mechanics}\ }\textbf
  {\bibinfo {volume} {623}},\ \bibinfo {pages} {329} (\bibinfo {year}
  {2009})}\BibitemShut {NoStop}%
\bibitem [{\citenamefont {Smith}\ and\ \citenamefont
  {Brutin}(2018)}]{smith2018wetting}%
  \BibitemOpen
  \bibfield  {author} {\bibinfo {author} {\bibfnamefont {F.}~\bibnamefont
  {Smith}}\ and\ \bibinfo {author} {\bibfnamefont {D.}~\bibnamefont {Brutin}},\
  }\href@noop {} {\bibfield  {journal} {\bibinfo  {journal} {Current Opinion in
  Colloid \& Interface science}\ }\textbf {\bibinfo {volume} {36}},\ \bibinfo
  {pages} {78} (\bibinfo {year} {2018})}\BibitemShut {NoStop}%
\bibitem [{\citenamefont {Chen}\ \emph {et~al.}(2016)\citenamefont {Chen},
  \citenamefont {Zhang}, \citenamefont {Zang},\ and\ \citenamefont
  {Shen}}]{chen2016blood}%
  \BibitemOpen
  \bibfield  {author} {\bibinfo {author} {\bibfnamefont {R.}~\bibnamefont
  {Chen}}, \bibinfo {author} {\bibfnamefont {L.}~\bibnamefont {Zhang}},
  \bibinfo {author} {\bibfnamefont {D.}~\bibnamefont {Zang}}, \ and\ \bibinfo
  {author} {\bibfnamefont {W.}~\bibnamefont {Shen}},\ }\href@noop {} {\bibfield
   {journal} {\bibinfo  {journal} {Advances in Colloid and Interface Science}\
  }\textbf {\bibinfo {volume} {231}},\ \bibinfo {pages} {1} (\bibinfo {year}
  {2016})}\BibitemShut {NoStop}%
\bibitem [{\citenamefont {Gao}\ \emph {et~al.}(2018)\citenamefont {Gao},
  \citenamefont {Huang},\ and\ \citenamefont {Zhao}}]{gao2018formation}%
  \BibitemOpen
  \bibfield  {author} {\bibinfo {author} {\bibfnamefont {M.}~\bibnamefont
  {Gao}}, \bibinfo {author} {\bibfnamefont {X.}~\bibnamefont {Huang}}, \ and\
  \bibinfo {author} {\bibfnamefont {Y.}~\bibnamefont {Zhao}},\ }\href@noop {}
  {\bibfield  {journal} {\bibinfo  {journal} {Science China Technological
  Sciences}\ }\textbf {\bibinfo {volume} {61}},\ \bibinfo {pages} {949}
  (\bibinfo {year} {2018})}\BibitemShut {NoStop}%
\bibitem [{\citenamefont {Gorr}\ \emph {et~al.}(2012)\citenamefont {Gorr},
  \citenamefont {Zueger},\ and\ \citenamefont {Barnard}}]{gorr2012lysozyme}%
  \BibitemOpen
  \bibfield  {author} {\bibinfo {author} {\bibfnamefont {H.~M.}\ \bibnamefont
  {Gorr}}, \bibinfo {author} {\bibfnamefont {J.~M.}\ \bibnamefont {Zueger}}, \
  and\ \bibinfo {author} {\bibfnamefont {J.~A.}\ \bibnamefont {Barnard}},\
  }\href@noop {} {\bibfield  {journal} {\bibinfo  {journal} {Langmuir}\
  }\textbf {\bibinfo {volume} {28}},\ \bibinfo {pages} {4039} (\bibinfo {year}
  {2012})}\BibitemShut {NoStop}%
\bibitem [{\citenamefont {Gorr}\ \emph {et~al.}(2013)\citenamefont {Gorr},
  \citenamefont {Zueger}, \citenamefont {McAdams},\ and\ \citenamefont
  {Barnard}}]{gorr2013salt}%
  \BibitemOpen
  \bibfield  {author} {\bibinfo {author} {\bibfnamefont {H.~M.}\ \bibnamefont
  {Gorr}}, \bibinfo {author} {\bibfnamefont {J.~M.}\ \bibnamefont {Zueger}},
  \bibinfo {author} {\bibfnamefont {D.~R.}\ \bibnamefont {McAdams}}, \ and\
  \bibinfo {author} {\bibfnamefont {J.~A.}\ \bibnamefont {Barnard}},\
  }\href@noop {} {\bibfield  {journal} {\bibinfo  {journal} {Colloids and
  Surfaces B}\ }\textbf {\bibinfo {volume} {103}},\ \bibinfo {pages} {59}
  (\bibinfo {year} {2013})}\BibitemShut {NoStop}%
\bibitem [{\citenamefont {Chen}\ and\ \citenamefont
  {Mohamed}(2010)}]{chen2010complex}%
  \BibitemOpen
  \bibfield  {author} {\bibinfo {author} {\bibfnamefont {G.}~\bibnamefont
  {Chen}}\ and\ \bibinfo {author} {\bibfnamefont {G.~J.}\ \bibnamefont
  {Mohamed}},\ }\href@noop {} {\bibfield  {journal} {\bibinfo  {journal}
  {European Physical Journal E}\ }\textbf {\bibinfo {volume} {33}},\ \bibinfo
  {pages} {19} (\bibinfo {year} {2010})}\BibitemShut {NoStop}%
\bibitem [{\citenamefont {Annarelli}\ \emph {et~al.}(2001)\citenamefont
  {Annarelli}, \citenamefont {Fornazero}, \citenamefont {Bert},\ and\
  \citenamefont {Colombani}}]{annarelli2001crack}%
  \BibitemOpen
  \bibfield  {author} {\bibinfo {author} {\bibfnamefont {C.}~\bibnamefont
  {Annarelli}}, \bibinfo {author} {\bibfnamefont {J.}~\bibnamefont
  {Fornazero}}, \bibinfo {author} {\bibfnamefont {J.}~\bibnamefont {Bert}}, \
  and\ \bibinfo {author} {\bibfnamefont {J.}~\bibnamefont {Colombani}},\
  }\href@noop {} {\bibfield  {journal} {\bibinfo  {journal} {European Physical
  Journal E}\ }\textbf {\bibinfo {volume} {5}},\ \bibinfo {pages} {599}
  (\bibinfo {year} {2001})}\BibitemShut {NoStop}%
\bibitem [{\citenamefont {Carre{\'o}n}\ \emph
  {et~al.}(2018{\natexlab{a}})\citenamefont {Carre{\'o}n}, \citenamefont
  {R{\'\i}os-Ram{\'\i}rez}, \citenamefont {Moctezuma},\ and\ \citenamefont
  {Gonz{\'a}lez-Guti{\'e}rrez}}]{carreon2018texture}%
  \BibitemOpen
  \bibfield  {author} {\bibinfo {author} {\bibfnamefont {Y.~J.}\ \bibnamefont
  {Carre{\'o}n}}, \bibinfo {author} {\bibfnamefont {M.}~\bibnamefont
  {R{\'\i}os-Ram{\'\i}rez}}, \bibinfo {author} {\bibfnamefont {R.}~\bibnamefont
  {Moctezuma}}, \ and\ \bibinfo {author} {\bibfnamefont {J.}~\bibnamefont
  {Gonz{\'a}lez-Guti{\'e}rrez}},\ }\href@noop {} {\bibfield  {journal}
  {\bibinfo  {journal} {Scientific Reports}\ }\textbf {\bibinfo {volume} {8}},\
  \bibinfo {pages} {9580} (\bibinfo {year} {2018}{\natexlab{a}})}\BibitemShut
  {NoStop}%
\bibitem [{\citenamefont {Carre{\'o}n}\ \emph
  {et~al.}(2018{\natexlab{b}})\citenamefont {Carre{\'o}n}, \citenamefont
  {Gonz{\'a}lez-Guti{\'e}rrez}, \citenamefont {P{\'e}rez-Camacho},\ and\
  \citenamefont {Mercado-Uribe}}]{carreon2018patterns}%
  \BibitemOpen
  \bibfield  {author} {\bibinfo {author} {\bibfnamefont {Y.~J.}\ \bibnamefont
  {Carre{\'o}n}}, \bibinfo {author} {\bibfnamefont {J.}~\bibnamefont
  {Gonz{\'a}lez-Guti{\'e}rrez}}, \bibinfo {author} {\bibfnamefont
  {M.}~\bibnamefont {P{\'e}rez-Camacho}}, \ and\ \bibinfo {author}
  {\bibfnamefont {H.}~\bibnamefont {Mercado-Uribe}},\ }\href@noop {} {\bibfield
   {journal} {\bibinfo  {journal} {Colloids and Surfaces B}\ }\textbf {\bibinfo
  {volume} {161}},\ \bibinfo {pages} {103} (\bibinfo {year}
  {2018}{\natexlab{b}})}\BibitemShut {NoStop}%
\bibitem [{\citenamefont {Deegan}\ \emph {et~al.}(1997)\citenamefont {Deegan},
  \citenamefont {Bakajin}, \citenamefont {Dupont}, \citenamefont {Huber},
  \citenamefont {Nagel},\ and\ \citenamefont {Witten}}]{deegan1997capillary}%
  \BibitemOpen
  \bibfield  {author} {\bibinfo {author} {\bibfnamefont {R.~D.}\ \bibnamefont
  {Deegan}}, \bibinfo {author} {\bibfnamefont {O.}~\bibnamefont {Bakajin}},
  \bibinfo {author} {\bibfnamefont {T.~F.}\ \bibnamefont {Dupont}}, \bibinfo
  {author} {\bibfnamefont {G.}~\bibnamefont {Huber}}, \bibinfo {author}
  {\bibfnamefont {S.~R.}\ \bibnamefont {Nagel}}, \ and\ \bibinfo {author}
  {\bibfnamefont {T.~A.}\ \bibnamefont {Witten}},\ }\href@noop {} {\bibfield
  {journal} {\bibinfo  {journal} {Nature}\ }\textbf {\bibinfo {volume} {389}},\
  \bibinfo {pages} {827} (\bibinfo {year} {1997})}\BibitemShut {NoStop}%
\bibitem [{\citenamefont {Pal}\ \emph {et~al.}(2019{\natexlab{a}})\citenamefont
  {Pal}, \citenamefont {Gope}, \citenamefont {Kafle},\ and\ \citenamefont
  {Iannacchione}}]{pal2019phase}%
  \BibitemOpen
  \bibfield  {author} {\bibinfo {author} {\bibfnamefont {A.}~\bibnamefont
  {Pal}}, \bibinfo {author} {\bibfnamefont {A.}~\bibnamefont {Gope}}, \bibinfo
  {author} {\bibfnamefont {R.}~\bibnamefont {Kafle}}, \ and\ \bibinfo {author}
  {\bibfnamefont {G.~S.}\ \bibnamefont {Iannacchione}},\ }\href@noop {}
  {\bibfield  {journal} {\bibinfo  {journal} {MRS Communications}\ }\textbf
  {\bibinfo {volume} {9}},\ \bibinfo {pages} {150} (\bibinfo {year}
  {2019}{\natexlab{a}})}\BibitemShut {NoStop}%
\bibitem [{\citenamefont {Pal}\ \emph {et~al.}(2019{\natexlab{b}})\citenamefont
  {Pal}, \citenamefont {Gope},\ and\ \citenamefont
  {Iannacchione}}]{pal2019comparative}%
  \BibitemOpen
  \bibfield  {author} {\bibinfo {author} {\bibfnamefont {A.}~\bibnamefont
  {Pal}}, \bibinfo {author} {\bibfnamefont {A.}~\bibnamefont {Gope}}, \ and\
  \bibinfo {author} {\bibfnamefont {G.~S.}\ \bibnamefont {Iannacchione}},\
  }\href@noop {} {\bibfield  {journal} {\bibinfo  {journal} {MRS Advances}\
  }\textbf {\bibinfo {volume} {4}},\ \bibinfo {pages} {1309} (\bibinfo {year}
  {2019}{\natexlab{b}})}\BibitemShut {NoStop}%
\bibitem [{\citenamefont {Carter}\ and\ \citenamefont
  {Ho}(1994)}]{carter1994structure}%
  \BibitemOpen
  \bibfield  {author} {\bibinfo {author} {\bibfnamefont {D.~C.}\ \bibnamefont
  {Carter}}\ and\ \bibinfo {author} {\bibfnamefont {J.~X.}\ \bibnamefont
  {Ho}},\ }\href@noop {} {\bibfield  {journal} {\bibinfo  {journal} {Advances
  in Protein Chemistry}\ }\textbf {\bibinfo {volume} {45}},\ \bibinfo {pages}
  {153} (\bibinfo {year} {1994})}\BibitemShut {NoStop}%
\bibitem [{\citenamefont {Abr{\`a}moff}\ \emph {et~al.}(2004)\citenamefont
  {Abr{\`a}moff}, \citenamefont {Magalh{\~a}es},\ and\ \citenamefont
  {Ram}}]{abramoff2004image}%
  \BibitemOpen
  \bibfield  {author} {\bibinfo {author} {\bibfnamefont {M.~D.}\ \bibnamefont
  {Abr{\`a}moff}}, \bibinfo {author} {\bibfnamefont {P.~J.}\ \bibnamefont
  {Magalh{\~a}es}}, \ and\ \bibinfo {author} {\bibfnamefont {S.~J.}\
  \bibnamefont {Ram}},\ }\href@noop {} {\bibfield  {journal} {\bibinfo
  {journal} {Biophotonics International}\ }\textbf {\bibinfo {volume} {11}},\
  \bibinfo {pages} {36} (\bibinfo {year} {2004})}\BibitemShut {NoStop}%
\bibitem [{\citenamefont {Gorr}(2013)}]{gorr2013lysozyme}%
  \BibitemOpen
  \bibfield  {author} {\bibinfo {author} {\bibfnamefont {H.~M.}\ \bibnamefont
  {Gorr}},\ }\emph {\bibinfo {title} {Lysozyme pattern formation in evaporating
  droplets}},\ \href@noop {} {Ph.D. thesis},\ \bibinfo  {school} {University of
  Pittsburgh} (\bibinfo {year} {2013})\BibitemShut {NoStop}%
\bibitem [{\citenamefont {Preibisch}\ \emph {et~al.}(2009)\citenamefont
  {Preibisch}, \citenamefont {Saalfeld},\ and\ \citenamefont
  {Tomancak}}]{preibisch2009globally}%
  \BibitemOpen
  \bibfield  {author} {\bibinfo {author} {\bibfnamefont {S.}~\bibnamefont
  {Preibisch}}, \bibinfo {author} {\bibfnamefont {S.}~\bibnamefont {Saalfeld}},
  \ and\ \bibinfo {author} {\bibfnamefont {P.}~\bibnamefont {Tomancak}},\
  }\href@noop {} {\bibfield  {journal} {\bibinfo  {journal} {Bioinformatics}\
  }\textbf {\bibinfo {volume} {25}},\ \bibinfo {pages} {1463} (\bibinfo {year}
  {2009})}\BibitemShut {NoStop}%
\bibitem [{\citenamefont {Pal}\ \emph {et~al.}(2019{\natexlab{c}})\citenamefont
  {Pal}, \citenamefont {Gope},\ and\ \citenamefont
  {Iannacchione}}]{pal2019image}%
  \BibitemOpen
  \bibfield  {author} {\bibinfo {author} {\bibfnamefont {A.}~\bibnamefont
  {Pal}}, \bibinfo {author} {\bibfnamefont {A.}~\bibnamefont {Gope}}, \ and\
  \bibinfo {author} {\bibfnamefont {G.~S.}\ \bibnamefont {Iannacchione}},\ }in\
  \href@noop {} {\emph {\bibinfo {booktitle} {International Conference on
  Pattern Recognition and Machine Intelligence}}}\ (\bibinfo {organization}
  {Springer},\ \bibinfo {year} {2019})\ pp.\ \bibinfo {pages}
  {567--574}\BibitemShut {NoStop}%
\bibitem [{\citenamefont {Neda}\ \emph {et~al.}(2002)\citenamefont {Neda},
  \citenamefont {Jozsa}, \citenamefont {Ravasz} \emph
  {et~al.}}]{neda2002spiral}%
  \BibitemOpen
  \bibfield  {author} {\bibinfo {author} {\bibfnamefont {Z.}~\bibnamefont
  {Neda}}, \bibinfo {author} {\bibfnamefont {L.}~\bibnamefont {Jozsa}},
  \bibinfo {author} {\bibfnamefont {M.}~\bibnamefont {Ravasz}},  \emph
  {et~al.},\ }\href@noop {} {\bibfield  {journal} {\bibinfo  {journal}
  {Physical Review Letters}\ }\textbf {\bibinfo {volume} {88}},\ \bibinfo
  {pages} {095502} (\bibinfo {year} {2002})}\BibitemShut {NoStop}%
\bibitem [{\citenamefont {Sendova}\ and\ \citenamefont
  {Willis}(2003)}]{sendova2003spiral}%
  \BibitemOpen
  \bibfield  {author} {\bibinfo {author} {\bibfnamefont {M.}~\bibnamefont
  {Sendova}}\ and\ \bibinfo {author} {\bibfnamefont {K.}~\bibnamefont
  {Willis}},\ }\href@noop {} {\bibfield  {journal} {\bibinfo  {journal}
  {Applied Physics A}\ }\textbf {\bibinfo {volume} {76}},\ \bibinfo {pages}
  {957} (\bibinfo {year} {2003})}\BibitemShut {NoStop}%
\bibitem [{\citenamefont {Katzav}\ \emph {et~al.}(2007)\citenamefont {Katzav},
  \citenamefont {Adda-Bedia},\ and\ \citenamefont {Arias}}]{katzav2007theory}%
  \BibitemOpen
  \bibfield  {author} {\bibinfo {author} {\bibfnamefont {E.}~\bibnamefont
  {Katzav}}, \bibinfo {author} {\bibfnamefont {M.}~\bibnamefont {Adda-Bedia}},
  \ and\ \bibinfo {author} {\bibfnamefont {R.}~\bibnamefont {Arias}},\
  }\href@noop {} {\bibfield  {journal} {\bibinfo  {journal} {International
  Journal of Fracture}\ }\textbf {\bibinfo {volume} {143}},\ \bibinfo {pages}
  {245} (\bibinfo {year} {2007})}\BibitemShut {NoStop}%
\bibitem [{\citenamefont {Goehring}\ \emph {et~al.}(2015)\citenamefont
  {Goehring}, \citenamefont {Nakahara}, \citenamefont {Dutta}, \citenamefont
  {Kitsunezaki},\ and\ \citenamefont {Tarafdar}}]{goehring2015desiccation}%
  \BibitemOpen
  \bibfield  {author} {\bibinfo {author} {\bibfnamefont {L.}~\bibnamefont
  {Goehring}}, \bibinfo {author} {\bibfnamefont {A.}~\bibnamefont {Nakahara}},
  \bibinfo {author} {\bibfnamefont {T.}~\bibnamefont {Dutta}}, \bibinfo
  {author} {\bibfnamefont {S.}~\bibnamefont {Kitsunezaki}}, \ and\ \bibinfo
  {author} {\bibfnamefont {S.}~\bibnamefont {Tarafdar}},\ }\href@noop {} {\emph
  {\bibinfo {title} {Desiccation cracks and their patterns: Formation and
  Modelling in Science and Nature}}}\ (\bibinfo  {publisher} {John Wiley \&
  Sons},\ \bibinfo {year} {2015})\BibitemShut {NoStop}%
\bibitem [{\citenamefont {Sobac}\ and\ \citenamefont
  {Brutin}(2014)}]{sobac2014desiccation}%
  \BibitemOpen
  \bibfield  {author} {\bibinfo {author} {\bibfnamefont {B.}~\bibnamefont
  {Sobac}}\ and\ \bibinfo {author} {\bibfnamefont {D.}~\bibnamefont {Brutin}},\
  }\href@noop {} {\bibfield  {journal} {\bibinfo  {journal} {Colloids and
  Surfaces A}\ }\textbf {\bibinfo {volume} {448}},\ \bibinfo {pages} {34}
  (\bibinfo {year} {2014})}\BibitemShut {NoStop}%
\bibitem [{\citenamefont {Giorgiutti-Dauphin{\'e}}\ and\ \citenamefont
  {Pauchard}(2015)}]{giorgiutti2015striped}%
  \BibitemOpen
  \bibfield  {author} {\bibinfo {author} {\bibfnamefont {F.}~\bibnamefont
  {Giorgiutti-Dauphin{\'e}}}\ and\ \bibinfo {author} {\bibfnamefont
  {L.}~\bibnamefont {Pauchard}},\ }\href@noop {} {\bibfield  {journal}
  {\bibinfo  {journal} {Soft Matter}\ }\textbf {\bibinfo {volume} {11}},\
  \bibinfo {pages} {1397} (\bibinfo {year} {2015})}\BibitemShut {NoStop}%
\bibitem [{\citenamefont {Lazarus}\ and\ \citenamefont
  {Pauchard}(2011)}]{lazarus2011craquelures}%
  \BibitemOpen
  \bibfield  {author} {\bibinfo {author} {\bibfnamefont {V.}~\bibnamefont
  {Lazarus}}\ and\ \bibinfo {author} {\bibfnamefont {L.}~\bibnamefont
  {Pauchard}},\ }\href@noop {} {\bibfield  {journal} {\bibinfo  {journal} {Soft
  Matter}\ }\textbf {\bibinfo {volume} {7}},\ \bibinfo {pages} {2552} (\bibinfo
  {year} {2011})}\BibitemShut {NoStop}%
\bibitem [{\citenamefont {Dillard}\ \emph {et~al.}(1994)\citenamefont
  {Dillard}, \citenamefont {Hinkley}, \citenamefont {Johnson},\ and\
  \citenamefont {Clair}}]{dillard1994spiral}%
  \BibitemOpen
  \bibfield  {author} {\bibinfo {author} {\bibfnamefont {D.~A.}\ \bibnamefont
  {Dillard}}, \bibinfo {author} {\bibfnamefont {J.~A.}\ \bibnamefont
  {Hinkley}}, \bibinfo {author} {\bibfnamefont {W.~S.}\ \bibnamefont
  {Johnson}}, \ and\ \bibinfo {author} {\bibfnamefont {T.~L.~S.}\ \bibnamefont
  {Clair}},\ }\href@noop {} {\bibfield  {journal} {\bibinfo  {journal} {Journal
  of Adhesion}\ }\textbf {\bibinfo {volume} {44}},\ \bibinfo {pages} {51}
  (\bibinfo {year} {1994})}\BibitemShut {NoStop}%
\bibitem [{\citenamefont {Behnia}\ \emph {et~al.}(2017)\citenamefont {Behnia},
  \citenamefont {Buttlar},\ and\ \citenamefont {Reis}}]{behnia2017spiral}%
  \BibitemOpen
  \bibfield  {author} {\bibinfo {author} {\bibfnamefont {B.}~\bibnamefont
  {Behnia}}, \bibinfo {author} {\bibfnamefont {W.~G.}\ \bibnamefont {Buttlar}},
  \ and\ \bibinfo {author} {\bibfnamefont {H.}~\bibnamefont {Reis}},\
  }\href@noop {} {\bibfield  {journal} {\bibinfo  {journal} {Materials \&
  Design}\ }\textbf {\bibinfo {volume} {116}},\ \bibinfo {pages} {609}
  (\bibinfo {year} {2017})}\BibitemShut {NoStop}%
\end{thebibliography}%

\newpage
\section*{FIGURES}

\end{document}